# John Ward: Memoir of a Theoretical Physicist

Norman Dombey

Department of Physics and Astronomy

University of Sussex Brighton UK BN1 9QH

August 4 2020

**To the Memory of Kate Pyne and Freeman Dyson**

who encouraged me to write this memoir of John Ward but were not able to see it.

**Abstract**  A scientific biography of John Ward, who was responsible for the Ward Identity in quantum electrodynamics; the first detailed calculation of the quantum entanglement of two photons in electron-positron annihilation with Maurice Pryce; the prediction of neutral weak currents in electroweak theory with Sheldon Glashow and Abdus Salam, and many other major calculations in theoretical physics.

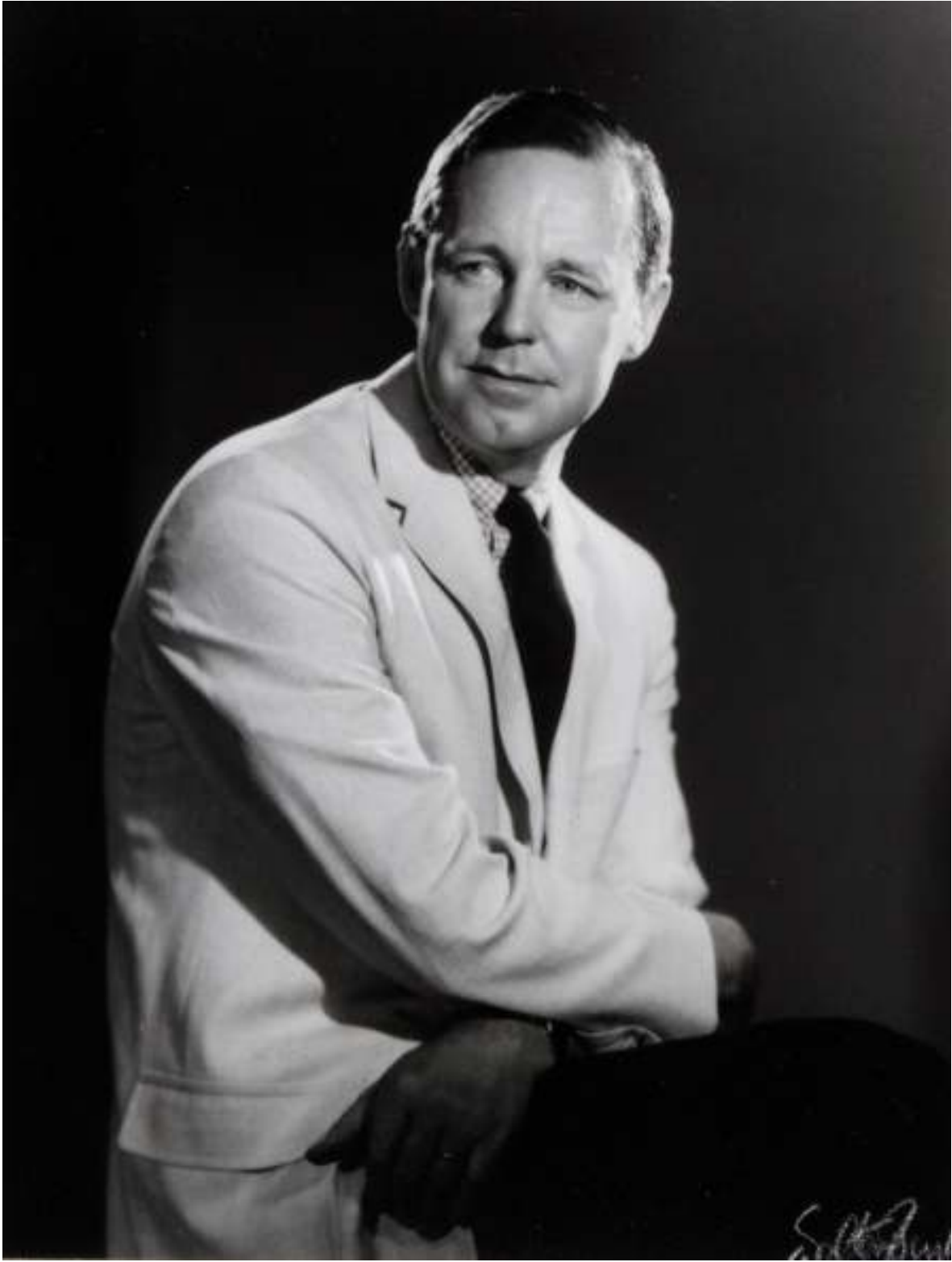

# SECTION HEADINGS

# 1) Introduction

John Clive Ward was a theoretical physicist who made important contributions to two of the principal subjects in twentieth century elementary particle physics, namely QED (quantum electrodynamics) and electroweak theory. He was an early proponent of the importance of gauge theories in quantum field theory and their use in showing the renormalisation of those theories: that is to remove apparent infinities in calculations. He showed that gauge invariance implies the equality of two seemingly different renormalised quantities in QED, a relationship now called the Ward Identity. This identity can be generalised to more general gauge theories in particle physics and remains a fundamental tool in these theories which dominate particle theory at the present time. He collaborated with Abdus Salam on the use of gauge theories in strong interactions and in electroweak theory. He also made significant contributions to statistical physics, in particular the two-dimensional Ising model, the electron gas, and second sound in liquid helium. In 1955 he was recruited by the UK Atomic Weapons Research Establishment at Aldermaston to head the Green Granite section of the theoretical group which had the task of rederiving the concept developed by Ulam and Teller in the United States. He spent the years from 1966 to his retirement in 1984 at Macquarie University in Australia.

Following his retirement at Macquarie he travelled widely and often came to Europe. I was interested in his work, especially his work at Aldermaston, and spoke to him several times about it. Section 8 contains an account of our conversations.

## 2) Early Years

Ward was born in East Ham in 1924 ar Barking just east of London. His parents were Joseph Ward, a civil servant in the Internal Revenue Department of the Treasury, and Winifred Ward, a schoolteacher. He showed early promise in mathematics and won a scholarship to Bishops Stortford College, a private school (in England it is known as a public school) where he boarded. As he says in his memoirs (27) 'the college was founded in the Victorian times to provide for the needs of the newly established Empire….in particular to create willing recruits for the Indian Civil Service. This was explicitly stated in the handbook that my father received". He found little intellectual stimulus in the school until he entered the science sixth form where he was able to ignore the "two science masters who seemed to know very little of science". The school however had a good science library which he read voraciously and so began his lifetime habit of self-education and of not interacting with his peers .As he says "being always a confirmed outsider, this habit became a lifelong resource". In 1941 he was entered for the entrance examination at Oxford when he was awarded a Postmastership (open scholarship) at Merton College.

## 3) Oxford and Quantum Entanglement

He studied mathematics in his first year and then transferred to engineering. He found very little in the course to interest him apart from a problem that A. M. Binnie was working on about the mechanical stresses due to gravity in symmetrical thin shells. Ward quickly realised that the solution only depended on the azimuthal dependence of the stresses and Ward and Binnie (1) published it in the Journal of the Royal Aeronautical Society. He obtained first class honours in Engineering and the following year a first in Mathematics.

Maurice Pryce was appointed to the Wykeham Professorship of Physics in 1946 and Ward became his first graduate student. Pryce suggested that Ward should work on QED. Dirac (1930) had shown that the two gamma rays resulting from the annihilation of a slow positron with an electron at rest would be polarised in perpendicular directions. Wheeler had just won a prize from the New York Academy of Science for a compilation of what was then known about the bound states of various numbers of positrons and electrons which he called polyelectrons. The simplest polyelectron was positronium, a hydrogen-like system with a positron in place of the proton. Wheeler (1946) suggested an experiment to test Dirac's result in which a slow positron beam incident on an electron at rest would create two photons

$$e^+ + e^- \to \gamma(1) + \gamma(2)$$

where the photon polarisations could be measured. Two groups were planning to carry out the experiment. So Pryce asked Ward to verify Wheeler's result that the numbers of photons polarised parallel and perpendicular to the scattering plane were different. Wheeler had calculated that the ratio of photons polarised perpendicular to the scattering plane to photons polarised parallel would have a maximum for scattering angle $74° \, 30'$.

Ward (27) later wrote that Pryce rejected his early attempt to duplicate Wheeler and pointed out that the state vector describing the two photons with polarisation labels $\alpha$ and $\beta$ travelling with momentum `k and –k in an angular momentum state J = 0 had to be the singlet state $|\alpha, \beta> - |\beta, \alpha>$. That was "my first lesson in quantum mechanics and in a sense my last since the rest is mere technique". More exactly Ward in his thesis (3) realised that the wave function of photon one and photon two had to be symmetric under interchange as they satisfied Bose-Einstein statistics and therefore the overall wave function is

$$|1, 2> = (|\alpha, \beta> - |\beta, \alpha>)(|k, -k> - |-k, k>) \qquad (1)$$

In particular the ground state wave function of positronium was known since it was just that of a hydrogen atom with proton mass equal to electron mass. Ward then proceeded to calculate the ratio of perpendicular and parallel polarisation states for two-photon annihilation of slow positrons incident on electrons using Heitler's (1944) time-dependent perturbation theory applied to radiation He showed that Wheeler had not used the correct two-photon wave function $|1, 2>$ above and therefore had neglected interference terms. Ward's result was that the maximum ratio occurred for 82º. [Figure 1]. Pryce and he then published a letter (2) in Nature including that result. Shortly afterwards, Snyder, Pasternack and Herrnbostel (1948) published a similar result in the Physical Review although Ward disagreed by a factor of 2 for the number of both perpendicular and parallel photon polarisations expected. Ward was correct.

Ward's full calculation (2),( 3) showed that the differential cross section for the process in terms of the scattering angles $\theta_1$ and $\theta_2$ of photons 1 and 2 and azimuthal angles $\varphi_1$ and $\varphi_2$ of their polarisations is proportional to

$$\frac{\{(1-\cos\theta_1)^3+2\}\{(1-\cos\theta_2)^3+2\}}{(2-\cos\theta_1)^3(2-\cos\theta_2)^3} -$$

$$\frac{\sin^2\theta_1 \sin^2\theta_2}{(2-\cos\theta_1)^2(2-\cos\theta_2)^2}\cos 2(\varphi_1-\varphi_2)$$

(2)

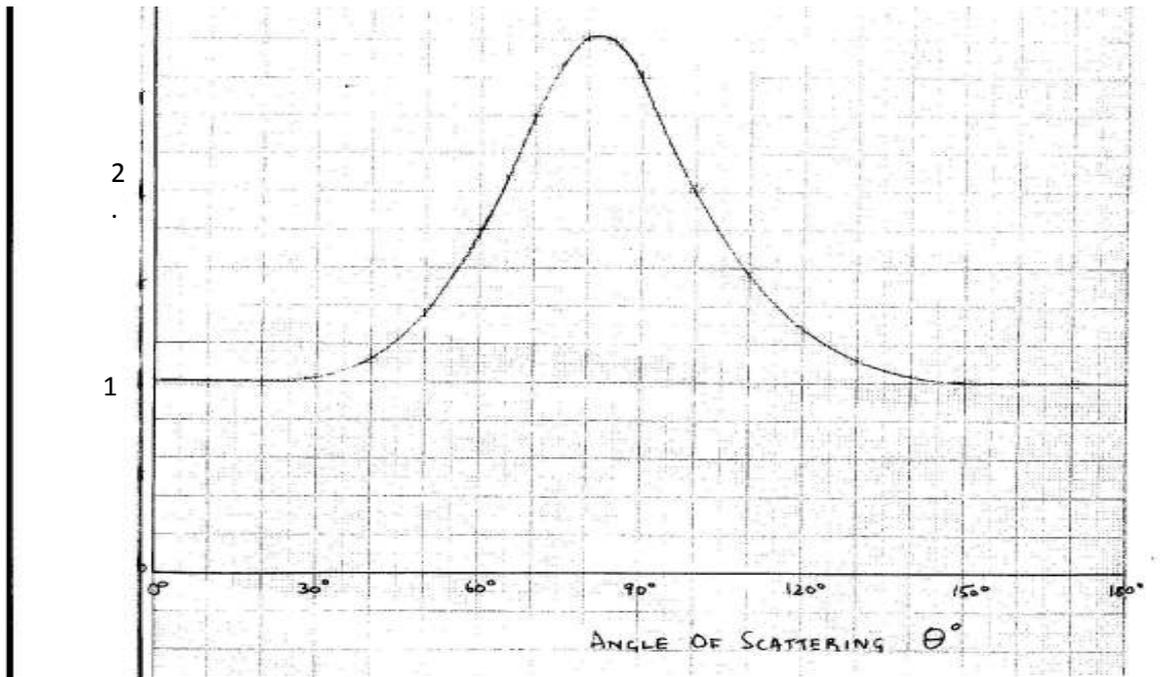

Figure 1

Ratio of perpendicular to parallel to photon polarisations as a function of scattering angle.

Hanna (1948) in Cambridge and Bleuler and Bradt (1948) in Purdue then published the results of their experiments. Taking account of the specific geometry used, the result of the Purdue experiment for the asymmetry ratio was

$$e^{\perp}/e_{||} = 1.94 \pm 0.37$$

compared with the theoretical value of

$$e^{\perp}/e_{||} = 1.7$$

This was very satisfactory although Hanna's result did not agree so well with the theory.

Einstein, Podolsky and Rosen (1935) discuss the results of the measurements of the physical quantities of two widely separated systems which can nevertheless be correlated. This is clearly the case here since the determination of the polarisation of one the photons gives information on the other's polarisation, even though the measurements can take place an arbitrary distance away. This is now referred to as quantum entanglement and it follows from Ward's two-particle wave function (Eq. 1). Dalitz many years later wrote in a letter to Ward that he "would regard the major achievement in your work [at Oxford] was the derivation of the two-photon wave function. ...I now know that it is the wave function for total spin zero but that was not known in those days" (RHD5). The physics of quantum entangled states in recent years has brought together an interdisciplinary field of research involving quantum optics, quantum computing and the foundations of quantum theory. The structure of Eq. (2) with its $[A + B \cos2(\varphi_1 - \varphi_2)]$ azimuthal dependence of the photon polarisations is typical of quantum entanglement (Duarte 2019).

Ward's calculation showed that he was already in the top rank of theoretical physicists working on QED. He had corrected Wheeler's result which had won a a major prize; had found a mistake of a factor of two in the results of Snyder et al, and had published his results in Nature before completing his doctorate. Yet instead of submitting this calculation he decided to work further on QED where there had been exciting new developments.

**4) Quantum Electrodynamics and the Ward Identity**

1947 and 1948 were especially important for the study of QED. Lamb and Retherford (1947) showed experimentally that the spectrum of the hydrogen atom according to the Dirac equation was not completely correct: there was a small correction to the $2S_{1/2}$ level which removed its degeneracy with the $2P_{1/2}$ state. Additionally the electron magnetic moment had been measured very accurately (Foley and Kush, 1948) and had a small deviation from the predicted value according to the Dirac equation of one Bohr magneton. In the two experiments use had been made of the microwave techniques developed in the UK and US during the war.

In both cases the small corrections to the results obtained from the Dirac equation were of order $\alpha/2\pi$ and thus should have been calculable using second order perturbation theory. But all the calculations which had been carried out had resulted in non-finite answers. For example the electron self-energy due to an electron emitting and reabsorbing a photon and the equivalent photon self-energy were both infinite.

The next year Julian Schwinger claimed that he could set up a framework which could explain these results and lay the foundations for a new fully relativistic field theory of electrons, positrons and photons. As he said in the introduction to his paper (Schwinger 1948) : "The unqualified success of quantum electrodynamics in applications involving the lowest order of perturbation theory indicates its essential validity for moderately relativistic particle energies. The objectionable aspects of quantum electrodynamics are encountered in virtual processes involving particles with ultra-relativistic energies. The two basic phenomena of this type are the polarisation of the vacuum and the self-energy of the electron.

The phrase 'polarisation of the vacuum' describes the modification of the properties of an electromagnetic field produced by its interaction with the charge fluctuations of the vacuum. … through the virtual creation and annihilation of electron-positron pairs by the electromagnetic field……The interaction between the electromagnetic field vacuum fluctuations and an electron….modifies the properties of the matter field and produces the self-energy of the electron…the vacuum polarization effects are equivalent to ascribing a proper mass to the photon….However the latter quantity must be zero in a proper gauge invariant theory." He goes on to show that although the self-energy of the electron calculated to order $\alpha/2\pi$ was divergent, when added to the bare or mechanical mass of the electron, the physical mass was obtained which could be assumed to be the finite physical mass $m$.

Similarly the vacuum polarisation process, although seemingly divergent, cannot contribute to the photon mass in a gauge-invariant theory but does change the electron charge from its bare charge $\hat{e}$ to its physical charge $e$. The transformation from the bare charge and mass of the initial QED Lagrangian to the physical $e$ and $m$ is called renormalisation and allows finite results to be obtained in calculations, at least to order $\alpha/2\pi$. Using this approach, expressions for the Lamb shift and the anomalous magnetic moment can be obtained which agree very well with experiment.

In addition to gauge invariance and renormalisation, Schwinger had to formulate the equal-time commutation relations of non-relativistic quantum field theory in covariant terms. This was a substantial task involving commutation relations of physical operators defined on space-like surfaces. But this was precisely the sort of problem that Ward enjoyed. In Part II of his thesis he attempted to generalise Schwinger's result on the electron self-energy in an external electromagnetic field to first order in $\alpha/2\pi$ to all orders. In this he was successful although he warned (3) that "it must nevertheless be repeated, and the forthcoming pages will underline this fact, that these results have only a rather formal interest. The detailed calculation of particular processes remains as difficult as ever."

This problem was extremely difficult and one in which Pryce could not be of much help. Appendix I gives a flavour of Ward's calculation. But Ward persevered and in April 1949 he submitted his thesis (3): Part I was his calculation of the angular correlation of polarisation states in $e^- + e^+ \rightarrow \gamma + \gamma$ while Part II was on his generalisation of Schwinger's results. The whole thesis ran to only 46 pages.

Rudolf Peierls from Birmingham was external examiner for Ward's thesis together with Jack de Wet as internal. Although Ward thought they gave him a hard time at his oral examination[1], the examiners are clear (Peierls and de Wet, 1949 ): the report on part I states that "This shows the author's competence in using methods of modern theory for a complex problem on which wrong statements had previously been published by distinguished people" and on part II "While in a subject of such rapid development and on which many experienced people are working intensely one cannot expect a D. Phil. candidate to make a major and lasting contribution, the candidate has succeeded in sorting out the new method from other scanty published information, and has succeeded in putting it in a form not given by Schwinger, which offers considerable advantage".

Pryce summarised Ward's doctoral work at Oxford thus: "Ward made seminal contributions to field theory even while still a student, but he was so brief in his explanations that people found it difficult to grasp his brilliant ideas". (Elliott and Sandars, 2005). Pryce considered that he had been one of his two most successful students at Oxford. and arranged for him to stay on for another two years with a DSIR[2] grant.

Ward began his postdoctoral career at Oxford in October 1949 after important new results in QED had been published: namely Richard Feynman (1949) had reproduced Schwinger's results but in a more intuitive way using a diagrammatic technique and Freeman Dyson (1949a) and (1949b) had shown that Feynman's and Schwinger's results were equivalent. Furthermore Dyson (1949b) conjectured that all the divergences of QED were contained in the mass and charge renormalisation to all orders of α/2π. Ward continued to work on the divergences in QED and how to remove them. Dyson had improved Schwinger's renormalisation technique by including the divergences in multiplicative constants, rather than removing them by subtraction. Then in terms of the renormalised fields $\psi(x)$ and $A_\mu(x)$, charge e and mass m can take their experimental values and Dyson conjectured that all orders of the perturbation series in terms of the renormalised charge and mass are finite. The renormalised electron and photon fields are obtained from the original bare electron spinor fields $\hat{\psi}(x)$ and photon fields $\hat{A}_\mu(x)$ by

$$\psi(x) = Z_2^{-1/2}\, \hat{\psi}(x) \tag{3a}$$

$$A_\mu(x) = Z_3^{-1/2} \hat{A}_\mu(x) \tag{3b}$$

where $Z_2$ and $Z_3$ are divergent quantities corresponding to the Feynman diagrams resulting from the photon and electron self-energies and are defined by the appropriate integrals in momentum space. They can be evaluated using a high momentum cutoff. $Z_1$ is a similar divergent quantity arising from the radiative corrections to the γee vertex where γ refers to a photon and e refers to an electron.

---
[1] In his memoirs Ward writes that Peierls "declared the thesis unworthy of acceptance". The written record gives a very different account.of Peierls' views.
[2] Department of Scientific and Industrial Research

Dyson (1949b) then showed that the relation between the bare electron charge $\hat{e}$ and the physical renormalised charge $e$ is

$$e = Z_1^{-1}\, Z_2\, Z_3^{1/2}\, \hat{e} \tag{4}$$

Adding one-loop processes: that is adding the electron self-energy term, the vacuum polarisation correction to the photon propagator and the correction to the vertex term due to a single photon being exchanged between the initial and final electrons allows the divergent quantities $Z_i$ to be calculated to order α/2π. The $Z_i$ are needed to obtain finite results when the renormalised fields are used. Dyson conjectured that $Z_1 = Z_2$ to all orders as it does in first order. This result would lead to a simplification of the proof of the renormalisability of higher order terms. Ward set out to prove this result to all orders.

He considered that gauge invariance

$$\widehat{\psi} \rightarrow e^{-i\hat{e}\widehat{\Lambda}} \widehat{\psi} \quad \widehat{A}_\mu \rightarrow \widehat{A}_\mu + \partial_\mu \widehat{\Lambda} \tag{5a}$$

must apply to the initial Lagrangian involving the bare operators $\widehat{\psi}$ and $\widehat{A}_\mu$ as the bare photon field described a massless particle. It should also apply to the renormalised operators $\psi, A_\mu$ in the renormalised Lagrangian since the mass of the physical photon is zero: i.e. the renormalised Lagrangian is invariant under

$$\psi \rightarrow e^{-ie\Lambda} \psi \quad A_\mu \rightarrow A_\mu + \partial_\mu \Lambda \tag{5b}$$

For Eqs. (5a) and (5b) to be simultaneously true we must have equal phases

$$e \Lambda = \hat{e} \widehat{\Lambda} \tag{6}$$

Using Eq. (3b) we get $\Lambda = Z_3^{-\frac{1}{2}} \widehat{\Lambda}$ and substituting Eq. (4) into Eq. (6) we obtain

$$e \Lambda = Z_1^{-1} Z_2 Z_3^{1/2} \hat{e} Z_3^{-1/2} \widehat{\Lambda} = \hat{e} \widehat{\Lambda}$$

so

$$Z_1 = Z_2 \tag{7}$$

This is now known as the Ward identity and is true to all orders in $\alpha/2\pi$. The use of the Ward identity reduces higher order self-energy terms which contain so-called overlapping divergences --- these are difficult to renormalise --- to vertex terms which are free of overlapping divergencies and can be renormalised. Dyson's QED renormalisation programme can therefore be carried out to all orders in $\alpha/2\pi$.

Ward's paper (6) entitled "An Identity in Quantum Electrodynamics" was written in his normal style: it had just one reference (to Dyson), eight equations and it took up less than half a page of the Physical Review.

Dyson (2016) told me more than 60 years later that: "the Ward papers on overlapping divergences demonstrated the deep connection between gauge invariance and renormalisability, which was another major step on the road to the standard model: the modern gauge theory of weak, electromagnetic and strong interactions. Ward did not make QED, but he transformed QED so that it fitted into the context of modern gauge field theory". Subsequently his work was generalised to more general gauge theories by Takahashi (1957) ,Taylor (1971) and Slavnov (1972).

## 5) Salam and Gauge Theories

Abdus Salam from Pakistan and Richard Dalitz from Australia were doctoral students of Nicholas Kemmer at Cambridge when Ward was a doctoral student at Oxford and both became lifelong friends. Dalitz (1947) did an independent calculation of the Hanna experiment which agreed with Ward while Salam (1952) wrote a Ph. D. on the renormalisation of QED including the higher order overlapping terms in perturbation theory. Ward went on to the Institute of Advanced Studies in Princeton where Robert Oppenheimer was Director and Dyson was a faculty member to continue his work on renormalisation; so did Salam when his thesis was completed at Cambridge.

Salam returned to the UK and after a spell back in Cambridge was appointed Professor of Theoretical Physics at Imperial College. while Ward stayed in the US getting temporary post-doctoral appointments for several years. They often corresponded. Weak interactions, in particular the properties of the neutrino and parity violation, became the new hot topic in the mid-1950s and Salam contributed to the subject with his work on the two-component massless neutrino and $\gamma_5$ -invariance (Salam 1957). Salam's student Ronald Shaw (1954) had written a thesis about what would happen if charged massless photons existed as well as a neutral massless photon and formed an isovector triplet of particles $\mathbf{A_\mu}$ coupled to the isovector current $i\,\bar{\psi}\gamma_\mu\,\tau\,\psi$ where now $\psi$ is a 2-component fermion spinor such as the nucleon spinor $\begin{pmatrix} p \\ n \end{pmatrix}$ and $\tau$ is the isovector formed of Pauli spin matrices . So instead of Eq. (5b) the equations of gauge invariance are now

$$\psi \rightarrow e^{-i\,\tau\cdot\Lambda}\psi \qquad \mathbf{A_\mu} \rightarrow \mathbf{A_\mu} + \partial_\mu\Lambda + \tau \times \Lambda \qquad (8)$$

and $\Lambda$ is an arbitrary isovector.

Shaw did not publish his work which was unfortunate because essentially the same analysis was published a short time later (Yang and Mills, 1954) emphasising that a theory of this type with a conserved isotopic spin current and the isovector gauge invariance of Eq.(8) would lead to the conservation of isotopic spin. Mathematicians would call the isotopic invariance of the theory based on the Pauli spin matrices $\tau_x, \tau_y, \tau_z$ the Lie group SU(2) generated by the Lie algebra composed of $\tau_x, \tau_y, \tau_z$.

The work of Yang, Mills and Shaw initiated a revolution in theoretical particle physics. Jun John Sakurai (1960) assumed that SU(2) could be applied to strong interactions and used it to describe the properties of the newly-discovered spin-1 particles $\rho^+, \rho^0, \rho^-$ which thus formed an isotopic vector: they were assumed to be the the massless gauge particles of Yang, Mills and Shaw but somehow had acquired mass. In electromagnetic and weak interactions Schwinger (1957) and then his student Sheldon Glashow (1961) thought that they could bring together QED and weak interactions by basing an electroweak theory on Yang, Mills and Shaw, where the gauge theory was based on SU(2) for weak interactions while keeping QED invariant under the one-dimensional unitary group U(1). If there were a common electroweak coupling the new electroweak spin-1 particles $W^+$, $W^0$, $W^-$ would have the very high mass of about 90 GeV: this was determined by the ratio of electric charge $e$ to $G^{1/2}$ where $G$ is the Fermi coupling of nuclear β-decay.

Ward joined Salam in an attempt to investigate gauge theories based on Lie groups applied to both strong interactions and electroweak interactions. They started with electroweak theory in 1959 with a paper 'Weak and Electromagnetic Interactions' (15) .This was modelled on Yang, Mills and Shaw where the basic isovector triplet now referred simply to ($W^+$, γ , $W^-$) where $W^+$ and $W^-$ mediated weak interactions and γ was a photon. But there was no explanation of how weak interactions violated parity while electromagnetic conserved it. Nor was there a reason for the large mass difference between the Ws and the photon.

In 1961 they published three papers on a gauge theory of strong interactions which included both isotopic spin and strangeness. In the first paper (19) they start with the σ-model of Gell-Mann and Levy (1960) but include the K as well as the π and σ . They then published two papers in which they followed Sakata (1956) by taking the basic 'nucleon' triplet as (p , n, Λ) but varied the Lie group: the special unitary group SU(3) in three dimensions (20) and the symplectic group in four dimensions Sp(4) were considered (21). The vector mesons in the theories were then determined by the gauge group: octets arise naturally in SU(3) and decuplets in Sp(4). The octet of vector mesons would consist of the isovector particles ($ρ^+$, $ρ^0$, $ρ^-$); and isoscalar $ω^0$ (as Sakurai had also assumed) and in addition four strange vector mesons with the same strangeness and isospins of the kaons. New particles and resonances of varied spins, isospins and strangeness were being discovered frequently at the time so it was not yet possible to choose which model would fit the experimental data.

At about the same time Salam's student Ne'eman and also Gell-Mann were working on similar lines using SU(3). Ne'eman (1961) obtained the same octet of vector gauge particles as Salam and Ward and in addition put the neutron and proton in an SU(3) octet with the same quantum numbers as the vector meson octet. He was thus able to include the strange isotopic baryon triplet ($\Sigma^+$, $\Sigma^0$, $\Sigma^-$) and the strange isoscalar $\Lambda$ in the octet. Independently Gell-Mann in his Eightfold Way (1961) was more systematic in his use of the group SU(3) and its different representations: he introduced an octet of pseudoscalars consisting of an isovector ($\pi^+$, $\pi^0$, $\pi^-$), two isospin-$\frac{1}{2}$ kaon doublets and he predicted a new isoscalar pseudoscalar meson $\eta^0$ which was promptly found by Pevsner (1961). Gell-Mann (1962) also predicted that the spin-$\frac{3}{2}$ baryonic resonances such as those found in pion-nucleon scattering must be part of a decuplet in SU(3) but only 9 particles with strangeness 0, -1, and -2 were known. So he predicted the isoscalar, strangeness -3 particle $\Omega^-$ which was soon found by Barnes et al. (1964).

Salam and Ward's attention returned to electroweak theory in 1964. They took their 1959 work but instead of the triplet ($W^+$, $\gamma$, $W^-$) they took an approach where the triplet ($W^+$, $W^0$, $W^-$) represented the vector mesons of the weak interactions while the field $A_0$ represented the photon. Then in order to have a single electroweak theory they allowed the field of the neutral weak vector meson $W^0$ to mix with the field $A_0$ obtaining

$$A = A_0 \cos\theta + W^0 \sin\theta \qquad Z = -A_0 \sin\theta + W^0 \cos\theta \qquad (9)$$

with electroweak mixing angle $\theta$. [I'm using a more modern notation than Salam and Ward used] where now Z is a new vector meson which becomes the quantum for a new electrically neutral weak interaction while A is the photon field after mixing. So this model based on the gauge group SU(2) x U(1) predicts a neutral weak current and the relationship between the electromagnetic and weak charges is

$$e = g \sin\theta \qquad (10)$$

where g is the coupling constant for the $\bar{\nu}\, e^- W^+$ and the $\bar{\nu}\, \mu^- W^+$ vertices $\bar{\nu}$ is an anti-neutrino and $e^-$ and $\mu^-$ are an electron and muon respectively[3]. Thus the Fermi coupling $G$ for muon decay into an electron, neutrino and anti-neutrino due to heavy $W^+$ exchange is given by

$$\frac{G}{\sqrt{2}} = \frac{g^2}{8 M_W^{\,2}} \qquad (11)$$

so using Eqs. (10) and (11) the W mass is given by

$$M_W = e/(2^{\frac{5}{4}} \sin\theta \, G^{\frac{1}{2}}) = (37.3/\sin\theta) \text{ GeV} \qquad (12)$$

particle physics. Salam considered that this paper, together with his earlier work on the two-component massless neutrino and $\gamma_5$ invariance, his most important achievements in physics and gave them as examples in his letter to the Chairman of the Nobel Committee describing his work on weak interactions (Salam 1969). He stated correctly in that letter that the paper predicted the existence of neutral weak currents. Yet the paper should not have been published because it only reproduced the results three years earlier of Glashow (1961). Furthermore neither of the crucial equations (10) nor (11) appear explicitly in the paper so that the authors do not note that the W mass must be at least 37.3 GeV and would be determined once $\sin\theta$ was measured.

## 6) An Itinerant Physicist and Statistical Physics

From 1951 to 1966, Ward was at the Institute for Advanced Study (1951–52, 1954–55, 1960–61); Bell Laboratories (1952–54); University of Adelaide (1954); University of Maryland (1956–57); University of Miami (1957–59); Carnegie Institute of Technology (1959–60), and Johns Hopkins University (1961–66). He was thus in the United States except for 1955 when he was back in the UK (see Section 8) and in Adelaide where he had accepted a position at the University but left after a few weeks to go back to the Institute of Advanced Study at Princeton.

---

[3] Salam and Ward assumed that the coupling of the W to the electron and neutrino was the same as its coupling to the muon and neutrino.

His work on QED and gauge theories is discussed above, yet he worked on various aspects of many body theory and statistical physics as well. As I haven't worked in this area I will rely on Ward's colleagues' recollections. His former student and colleague Francisco Duarte who edited his memoirs commented "In addition to these Herculean contributions [on QED and gauge theories)] John collaborated with well-known mathematicians and theoreticians .....producing a series of brilliant papers on: the Ising Model (12), quantum solid-state physics (11), quantum statistics (14), and Fermion theory (17) (Duarte 2009). Roger Elliott who knew him as a graduate student, told me that Ward had some interaction at Oxford with the low-temperature group. He went on to write papers (10) , (11) about second sound in liquid helium with Wilks. "subsequently Ward showed the enormous range of his versatility by making at least two significant contributions in the area of statistical physics. While working with Montroll in Maryland he took part in a discussion (22) of Onsager's remarkable solution to the two-dimensional Ising model. John proposed that the results might be obtained in a more direct way using combinatorial mathematics and in collaboration with Mark Kac (12) demonstrated this with another remarkable mathematical tour de force. Subsequent papers showed that it was a complete solution but disappointingly did not help with the many doomed attempts to extend it to the three-dimensional case. In another occasion, again working with Montroll, he showed that the state of the electron gas which had been extensively studied at absolute zero could be extended by what was effectively a form of to Debye-Huckel theory. In this work (14) he also was probably the first to use the periodicity of properties in inverse temperature. Either of these examples would have ensured him a enviable position amongst statistical physicists but they remained for him a side interest" (Elliott 2017).

## 7) Macquarie

Throughout his itinerant years, Ward was uneasy with American physics graduate schools. According to his memoirs they were established so that "physicists would be created by a production line of suitable professors, financed by the splendid generosity of government institutions. The same students could then also be used as cheap instructors for undergraduate courses, leaving the professors with the help of his contract, to spend his time directing research, and publishing more papers. This resulted in an absurd inflation of theoretical physics in particular, aided and abetted by publishers of innumerable semi-fraudulent 'science journals'." At some time in the future the graduate student would have to join "an ever-growing army of unfortunate ex-graduate students, now professors, and be competing for ideas" as best he could. He thought that this was quite impossible.

Dalitz and Salam[4] often urged him to return to the UK when vacancies arose. The search for publishable papers was not as extreme in Britain as in the US but it was probably just a matter of time before Britain caught up, so he did not take them up. He did, however, allow his name to go forward to be a candidate for a fellowship of the Royal Society and he was elected a Fellow in 1965. This helped him to look for a Professorship outside the US and Britain.

---

[4] Salam's correspondence with Ward is held in the Salam archive of the Salam International Centre for Theoretical Physics in Trieste, Italy; Dalitz' correspondence is held in the Radcliffe Science Library, Oxford, UK. Unfortunately at the present time neither is publicly available.

He saw an advertisement in Nature for a Professor of Mathematics in Wellington, New Zealand. He applied and was offered the job. Soon after he arrived in Wellington, he went to Sydney and met an old friend there -- Freddie Chong -- who had just been appointed Professor of Mathematics at Macquarie University. Chong persuaded Ward to transfer to Macquarie which was a new university.

Unfortunately for Chong and Ward, Macquarie in 1966 was run by non-scientists and most students were teacher trainees studying for a B. A. degree. Physics was a new subject and Ward was in charge of developing courses and of recruiting faculty. He enjoyed undergraduate teaching and chose 'The Feynman Lectures on Physics' as the key text supplemented by more conventional physics text books. He found that he much preferred this sort of teaching to the supervision of endless graduate students in the search for 'golden eggs' as he termed research achievements. He also introduced a Master's course for local physics teachers, also based on the Feynman Lectures which was very successful. He supported Duarte and the science students who campaigned for a B. Sc. degree to be awarded as well as the B.A. This goal was finally attained in 1979.

Ward stayed at Macquarie from 1966 until his retirement in 1984. Yet he did not publish in a recognised physics journal while he was there. His interest in particle theory continued but he insisted that the publication of any new paper of his must meet the standards of his previous contributions to physics. Ward was probably reacting to Salam's regular practice of publishing the latest thought in his head. For example, according to Kibble (1998), Salam published 17 papers in 1975. After all no one took account of papers which were soon forgotten, whereas papers which were noticed added to one's reputation. Ward stated in his memoirs "He [Salam] and I were old friends, despite the fact that our temperaments were directly opposite. He would publish anything and hope for the best. I would not normally publish unless I was sure of the product."

According to Duarte (2009) "at Macquarie he [Ward] became known for his forceful defence of science, high academic standards, and for his uncompromising honesty. In this regard, he openly and vigorously supported the student science reform movement that permanently changed the degree structure of the university. This transformative innovation strengthened significantly the structure of the sciences at Macquarie".

**8) Fermi, Ulam and Teller**

After the second world war QED and quantum field theory became the principal research interests of theoretical physicists (see Section 4 above). Robert Oppenheimer moved into civilian life as the Director of the Institute of Advanced Study at Princeton and set up conferences in 1947 and 1948 for the National Academy of Sciences on the subject of "Foundations of Quantum Mechanics" and in particular on progress in QED in the light of the new experiments on the Lamb shift and the anomalous magnetic moment of the electron. The first was held from June 2 to June 4, 1947 in Shelter Island, New York, and the second in the Pocono mountains in Pennsylvania from March 30 to April 2 1948. The guest list in both was restricted to those who had made recent progress in the fields of QED and particle theory (Schweber 1994). The discussion leaders at Shelter Island were Kramers, Oppenheimer and Weisskopf. Kramers (1938) was the originator of the concept of mass renormalisation of the electron when describing the classical interaction of the electron with its radiation field; Oppenheimer (1930) attempted to describe the relativistic interaction of the electron with its radiation field, and Weisskopf (1939) had shown that in QED the self-energy of the electron due to its interaction with its radiation field was only logarithmically infinite. According to Schweber, Schwinger and Feynman gave accounts of their approach to QED at Pocono and there were presentations of the calculation of the Lamb shift by Bethe, of the theory of hyperfine structure by Teller, of the neutron-electron interaction by Fermi, and of an analysis of meson experiments by Serber.

These conferences were remarkable for a reason not directly related to the foundations of quantum mechanics and QED, In Project Y: the Los Alamos Story Part I, Hawkins (1983) shows that apart from Kramers and Schwinger, all those mentioned above played a major role at Los Alamos in the Manhattan Project during the war: Oppenheimer was Scientific Director of the whole Project; Bethe was Head of the Theoretical Division, while Fermi was Head of F Division which included work on possible thermonuclear weapons. Weisskopf, Serber and Feynman were Deputy Division Heads in the Theoretical Division and Teller was a Deputy Division Head in F-Division.

Following the war and the deepening split between the United States and the Soviet Union, there was much discussion between those physicists led by Oppenheimer who considered that fission weapons were sufficient in any future war and those led by Edward Teller who considered that thermonuclear weapons based on the fusion of light elements were needed. Fission weapons typically had an explosive yield measured in kilotons of TNT while the yield of a fusion weapon would be measured in megatons. Two events in 1949/1950 settled the matter. The USSR detonated a fission weapon on 29 August 1949 and then Klaus Fuchs, who had worked in the Theoretical Division as a member of the British Mission at Los Alamos during the war, confessed on 27 January 1950 that he had spied for the Soviet Union and had provided details of the weapon designs developed in the Manhattan Project, see Close (2019), President Truman announced on 31 January 1950 that he had directed the US Atomic Energy Commission "to continue its work on all forms of atomic weapons including the so-called hydrogen or superbomb" Rhodes (1995) gives a full account of the debate between Oppenheimer and Teller, and Truman's eventual decision in 'Dark Sun'.

The major physics problem involved in a hydrogen bomb is that the energy release arises from the fusion of deuterium and tritium:

$$D + D \rightarrow {}^3He + n + 3.3 \text{ MeV}$$

$$D + D \rightarrow {}^3H + p + 4.0 \text{ MeV}$$

$$D + T \rightarrow {}^4He + n + 17.6 \text{ MeV}$$

but D and T are positively charged so in order to fuse, they have to surmount the Coulomb barrier to get within a few nuclear radii $r_D$ of each other. The height of the Coulomb barrier when the two nuclei encounter each other is about $\alpha / r_D$ where $\alpha$ is the fine structure constant. So the barrier height is many keV. Deuterium or tritium must therefore be heated to a temperature of tens of millions of degrees to have a chance of extracting explosive energy from fusion. This can only be achieved by exploding a fission weapon. Furthermore at such high temperatures all the matter completely ionises and every electron and positive ion will radiate energy away.

In August 1945 Enrico Fermi gave a series of lectures at Los Alamos on the current state of work on a possible thermonuclear weapon which was then known as the Super. Members of the British Mission still at Los Alamos were present. Philip Moon of the University of Birmingham took notes which he passed on to James Chadwick, the Mission leader. They were subsequently transferred to London and G P Thomson received a copy which now resides with his papers in Trinity College library in Cambridge. Moon (1945) writes that Fermi considered the balance between the rate of energy produced by deuteron-deuteron [D-D] fusion and the rate of energy loss due to radiation. Fermi continues "It will be necessary to consider transfer of energy to ions, electrons and radiation. ...W is the energy released by the reaction between two nuclei...if the system of electrons and ions is in thermal equilibrium . the acceleration of electrons due to collisions with ions causes a transfer of energy to the radiation field at a rate (r)..... .the reaction will progress if W > r .....Since both W and r are proportional to $n^2$......[ $n$ is the concentration of nuclei per cubic centimetre] the [critical temperature] $θ_{crit}$ [defined as the temperature at which r = W] is independent of concentration.

In 1992 after the breakup of the Soviet Union, the nuclear weapon institute at Sarov (Arzamas-16 in Soviet times and the Soviet equivalent of Los Alamos) published a history of the Soviet H-bomb programme (Goncharov and Maksimenko 2008) . It included another copy of Fermi's lectures, presumably written by Fuchs and given to his Soviet contact.. Fuchs' version goes straight to the point, as we will see later. He writes that Fermi concluded in his summary of the same lecture "If thermal equilibrium between particles and radiation were established, it would be impossible [to] heat deuterium to required temperature. In actual fact there will be no thermal equilibrium".

So Fermi's conclusions were that to ignite the Super and obtain W > r it would be necessary to heat the D and T ions to a higher temperature than that of the radiation and that compression of the heavy hydrogen fuel made no difference.

Throughout 1950 at Los Alamos Ulam and Everett calculated the dynamics of deuterium plus various amounts of tritium at the temperature attainable using a fission weapon. Tritium produced more energy from a DT interaction than a corresponding DD reaction did. The question they set out to answer was "Did the temperature increase, hold its own, or diminish" (Ulam 1976). They found that unrealistically large amounts of tritium would be needed to maintain the temperature. With attainable amounts of tritium, the Super would fizzle.

On March 9 1951 everything changed. Ulam and Teller wrote the report 'On Heterocatalytic Detonations I; Hydrodynamic Lenses and Radiation Mirrors' (Ulam and Teller 1951) . This laid the foundation for modern H-bombs. To translate: catalytic means using a substance that increases the rate of a reaction without itself changing; autocatalytic means that the reaction products themselves increase the rate of the reaction, while heterocatalytic means that one reaction is driving another reaction: in this case fission reactions in the primary are initiating fusion reactions in the secondary. These two different types of reaction take place in separate physical spaces within a uranium container (Teller 1976). Hydrodynamic lenses and radiation mirrors refer to the container's capacity to direct the radiation from the primary to implode the secondary. On November 1 1952 there was a successful test of Ulam and Teller's ideas when a device, codenamed Mike, was exploded in the Marshall Islands yielding 10.4 megatons. Mike weighed 82 tons and was not a deliverable weapon but was proof that the Ulam-Teller [U-T] concept worked (Rhodes 1995).

In Britain there was a similar debate about whether to attempt to build a hydrogen bomb. The UK had first tested an A-bomb in October 1952 but after the successful Mike test in 1952 and especially the Soviet announcement of a successful H-bomb test yielding 400 kilotons [named Joe-4 in the west] in August 1953 (Holloway 1994) , pressure was growing in military and political circles for Britain to build an H-bomb. Churchill was Prime Minister at the time and he was heavily influenced by his Scientific Advisor Lord Cherwell, who was also head of Oxford Physics Department. "Cherwell, in fact, had already been pressing Penney [the head of Britain's nuclear weapon programme] to start work on the superbomb" in 1953 (Penney and Macklen 1988) . The government decided in July 1954 to make the H-bomb and the Atomic Energy Authority came into existence in August (Arnold 2001).

Aldermaston in Berkshire then became the main site for the H-bomb work and William Cook joined Penney there as his deputy with responsibility for designing the H-bomb. Of course no one at Aldermaston knew what Ulam and Teller had done since the US McMahon Act forbade any transmission of nuclear information to other states.

John Ward says in his memoirs that "One day I received a letter, much to my surprise, from Francis Simon. He asked whether I was enjoying life at Bell Labs." He also suggested that Ward return to Oxford. Simon was head of the Oxford Low Temperature Group and Ward did not know him particularly well. Ward replied that he would like to return to Oxford and was then told by Simon that "he regretted that he did not actually have any positions at the moment".

That Simon did not have any position to offer Ward can be explained. Simon was close to Cherwell whom he had met in Berlin around 1920 : he was Jewish and was brought by Cherwell to Oxford in 1933 after Hitler took power. In his governmental role Cherwell in 1953 wanted to attract theoretical physicists who might be able to help Penney rediscover U-T. Who better than an Oxford graduate with a worldwide reputation in QED just like the American nuclear weapon physicists?[5] After all Dyson had even been invited to work on H-bombs by the U.S. Atomic Energy Commission even though he was a British citizen (Dyson 1992). But Ward could not then be offered a position on the British H-bomb programme as it had not yet been agreed by the government.

After the government's decision to proceed, Cook began work at Aldermaston in September 1954 and immediately advertised for theoretical physicists. Ward applied and according to his memoirs Cook told him that the matter was urgent and that his presence was very much desired. Ward returned to the UK and began work at Aldermaston in June 1955. A few days after he arrived Ward says in his memoirs that "there was a formal meeting, chaired by Penney, of about 20 senior staff. He declared that I would be in charge of *Green Granite*, the codename for the development of a U-T device. A few days later he was called to Penney's office with Keith Roberts, a theoretical physicist on Aldermaston's staff, and told what was known of U-T at Aldermaston, namely that the device involved two-stages with a fission primary and a fusion secondary, and that neutron shielding was involved. That was it. It was up to Ward, with Roberts' assistance, to rediscover what Ulam and Teller had worked out.

---

[5] In his memoirs, Ward says that Kramers had been impressed by Ward's work on both renormalisation and the Ising model and had written to Simon saying that Simon should try to attract him back to Britain. Dalitz searched for Kramers' letter and could not find it in Simons' papers (RHD2,RHD3,RHD4). I think it is more likely that Simon was writing on behalf of Cherwell and that Kramers had nothing to do with Simon's letter.

In the 1980s when the cold war was at its coldest, I was a member of the British Pugwash Group which tried to bring western and Soviet scientists together to reduce tensions. I knew several of the physicists who had worked on nuclear weapons in the US, Britain and the USSR. I was taught QED at Caltech by Feynman; Peierls, who with Fritsch wrote the original memorandum at Birmingham showing that a nuclear weapon was possible was also a member of Pugwash, and I had spent 1962-63 in Moscow on a British Council exchange where I had met many distinguished Soviet physicists who had worked in the Soviet nuclear programme. In 1989 I had invited Andrei Sakharov to Sussex where he was awarded an honorary degree. I knew that Ulam and Teller were the originators of thermonuclear weapons in the US and that Sakharov's ideas were responsible for Joe-4 in the USSR. I wondered who had rediscovered U-T in the UK. Everything about the British H-Bomb was still classified in the 1980s.

In 1991 I was reading 'US Nuclear Weapons: The Secret History' ( Hansen 1988) when I came across a footnote "To my amazement, when (in 1955) I reached Aldermaston (the British nuclear weapons development lab ....I was assigned the improbable job of uncovering the secret of the Ulam-Teller invention ...an act of genius far beyond the talents of the personnel at Aldermaston, a fact well known to both Cook and Penney".  Hansen claimed that Ward had  written a letter from Macquarie  to Mrs. Thatcher, then Prime Minister, claiming that he had succeeded in his task  and that his work should be recognised (see Appendix II)..

I had met Ward in Brookhaven in summer 1964 when  we were both visiting. We hadn't kept in touch.  Maurice Pryce visited my department at Sussex occasionally in the 1990s and I knew he had been Ward's supervisor. He confirmed that Ward had been employed at Aldermaston. So I wrote to Ward at Macquarie on April 30 1991  asking whether he ever came to Europe since I would like to discuss his time at Aldermaston. The previous year  I had asked former  senior Aldermaston officials at a seminar on  British post-war  nuclear policy who were the British equivalent of Ulam and Teller  and all insisted that everything was done collaboratively. No single person could claim authorship of the U-T concept in Britain (ND1).

Five months later I received a reply from Canada. Ward had retired and had left Macquarie . He said that he remembered me and that "Anyone who believes that radiation implosion could be invented by committee should be locked up.  Quite simply I was drafted in  to tell them how to do it, and Penney refused to listen. I think it was basically a fight between Cherwell  and Cook on the one hand, and Penney on the other, with me used as a pawn" (JCW1) .  He then made some remarks about the "criminal activities of MI6" and finished saying that  he would be in  Portugal in December.  In fact he delayed his visit to Portugal to April 1992 but I received a phone call  that autumn saying that he was in Calais and could I come to see him. So the following day I set out  by train and boat to Calais. We both recognised each other and he told me his Aldermaston story which was basically  what he wrote in his memoirs in 2004.

In April I went to see him again in Cascais a pleasant coastal town near Lisbon. Ward wanted to tell his story more widely and I had arranged via a colleague at Lisbon University for a journalist on Publico, a Lisbon newspaper, to interview him. I wrote an introductory piece in English which was translated as "A bomba de hidrogenio 40 anos depois" [40 years of hydrogen bombs] (Dombey 1992). When I arrived at Cascais I showed Ward my piece. I had written that "he [Ulam] was struck by the idea of the two-stage process and realized that the electromagnetic radiation from the primary explosion could be reflected by a heavy, naturally occurring metal such as uranium on to the secondary". Ward became very agitated. "Not reflected" he shouted. I thought for a bit and couldn't understand why not. Then I said tentatively "black-body radiation". He calmed down and nodded.

Ward told his story in the interview. It was reported in the Independent on Sunday (Cathcart 1992) and then the military historian Eric Grove and I followed it up in the London Review of Books in an article called 'Britain's Thermonuclear Bluff' (Dombey and Grove, 1992). Information about the development of Britain's H-Bomb was unavailable so it was useful for us to be able to refer to published sources. When Aldermaston's historian Lorna Arnold published her official account 'Britain and the H-Bomb' (Arnold 2001), Appendix 5 is devoted to Ward's claim to have reinvented U-T. Arnold inspected the written record for 1955: she confirms that Ward had been group leader of 'new devices' and that Roberts helped him even though "he [Ward] had, as he said, worked almost entirely alone." The theoretical physicists at Aldermaston numbered 82 scientists and produced 123 papers that year, none of which were by Ward. Roberts, however, had written a paper ' An Elementary Theory of Detonations' (Roberts 1955) after Ward had left Aldermaston which referred to Ward and must have included Ward's results. Note that Roberts follows Ulam and Teller in using the word 'detonation' in his title so his paper is presumably the British equivalent of U-T.

I managed to arrange for Ward to defy MI6 and come to Britain in 1993 where I introduced him to Lorna Arnold. He didn't say anything new but drew some sketches of bombs which he gave to her. In her book Arnold says that "it seemed clear from his answers and the sketch that the Ward concept, whatever its intrinsic value, had not been the basis of the various Grapple devices (a fact he could not have known without access to later British work)".

Kate Pyne, Lorna Arnold's successor, was interested in a further study of Ward's contribution and discussed it with me. She wrote to me in July 2008 (Pyne 2008) saying that "I really don't know what it was that Professor Ward claimed to have found or discovered all those years ago". Then on 20 June 2015 I received an email from her saying that "I don't wish to denigrate Lorna Arnold's memory, but I didn't agree with her about relegating Dr Ward to a relatively short Appendix in 'Britain and the H-bomb'. I'm sure a more appropriate tribute could be written so I'm glad to hear that you've contacted the RS." (Pyne 2015).

Kate died from a pulmonary embolism less than two weeks later. This rather long section is my attempt to fulfil her wishes. Kate's own discussion of Ward's contribution is contained in her posthumous thesis at Kings College London where Chapter 5 is called "In the Matter of John Clive Ward: Work on thermonuclear warheads in Britain between June and December 1955. [What progress was made in the programme on thermonuclear warheads between June and December of 1955 when John Ward worked at AWRE?]" (Pyne 2016). Unfortunately that thesis is kept at Aldermaston under lock and key. Kate clearly did find progress as a result of Ward's presence at Aldermaston. So I will give my answer to what he had discovered using readily accessible material and some elementary physics.

Richard Rhodes writes in 'Dark Sun' about the work showing that the Super did not work and the circumstances which led to Ulam and Teller's new concept. Freeman Dyson continues the story in his biographical memoir of Teller (Dyson 2007). "In 1950 electronic computers were able to simulate in a rough fashion the Classical Super design for a hydrogen bomb and showed that it did not work. George Gamow drew a famous cartoon of Teller trying to set fire to a wet piece of rock with a match. But to Teller the downfall of the Classical came as a liberation. For eight years his thoughts had been fixed on the Classical Super, which required deuterium to burn at low density, so that radiation could escape from the burning region and not come to thermal equilibrium with the matter [see Fermi's lecture notes discussed earlier].The idea was to achieve a runaway burn with temperature of the matter remaining much higher than the temperature of the radiation. The computers showed that runaway burn did not work. So Teller began to look seriously at the opposite situation, with deuterium at high density and the radiation trapped in thermal equilibrium with the matter. Teller found that ar high density, deuterium could burn well in thermal equilibrium with the matter.....Stanislaus Ulam at Los Alamos thought of a similar arrangement at the same time, and so the idea became known as the Teller-Ulam design."

From the title of the U-T paper and Dyson's note, it is possible to see that the crucial ideas of the U-T concept are (i) a physically-separated primary and secondary; (ii), thermal equilibrium between radiation and matter and (iii) high compression of the secondary. Teller had shown in his article in Encyclopedia Americana (Teller 1976) that the primary and secondary are inside a uranium-238 container. In the minutes of the meetings between US and UK physicists in 1958 after the UK had demonstrated that it could explode H-bombs, it was reported that the UK had tested two-stage radiation implosion weapons (USAEC 1958). Ward claimed that he had reinvented U-T and in particular radiation implosion. (Appendix II, (ii) and (v); JCW 2a).

Arnold reports that Penney was aware in September 1955 that a two-stage device should be used so Ward was not responsible for that idea. That compression of the deuterium (or deuterium plus tritium) fuel was necessary was known to him: in his memoirs he remembers that at an important meeting at Aldermaston where he had discussed his ideas he "explained how important it was to move the energy fast to the other end and emphasized the need for compression". He also wrote to Dalitz in May 1997 "After my resignation it was realised my ideas about radiation implosion, compression and subsequent heating were all correct" (JCW 2a). The need for high compression for ignition in inertial fusion experiments is demonstrated in 'The Physics of Inertial Fusion' where the compression necessary to ignite 1 mg of DT fuel is shown to be 1500 (Atzeni and Meyer-ter-Vehn 2004) . That compression helps ignition is not difficult to understand since compression will push the deuterium or tritium nucleus up the Coulomb potential barrier so less energy and thus a lower temperature is needed for fusion.

When Ward insisted that I must not say that the uranium container reflected the radiation from the primary on to the secondary and that black-body radiation was involved, it could only mean that he envisaged that the radiation was in equilibrium with the matter inside the container. Physics undergraduates are taught how to calculate electromagnetic radiation pressure and at room temperature it is negligible. But the Stefan-Boltzmann law shows that the relation between the radiation pressure $S$ and the equilibrium temperature T is

$$S = AT^4 \text{ W/cm}^2 \quad (12)$$

where $A$ is a constant  The temperature of an A-bomb is about $50 \times 10^6$ K so that the pressure arising from the radiation at that temperature in equilibrium would dwarf any pressure arising from matter. For example, the radiation pressure on the sides of a box at a temperature of 300 K is $20 \times 10^{-12}$ atmosphere. But raise the temperature to $50 \times 10^6$ K and the pressure increases to $2 \times 10^9$ atmospheres (Ford 2015).

Using radiation to compress the secondary to ignite the heavy hydrogen is called radiation implosion. Ward clearly was considering the situation where radiation and matter were in equilibrium when he was talking to me and that must have been the result of his work in 1955. Atzeni and Meyer-ter-Vehn state that the radiation flux on a target in a cavity of gold increases as $T^4$ when "multiple absorption and re-emission processes lead to a thermal distribution of photons in the cavity described by black body radiation" in inertial fusion[6]. The use of radiation to obtain extreme pressures and therefore extreme compression followed from the thermal equilibrium of radiation and matter that Ward introduced, just as Ulam and Teller had done four years earlier.

Ward's claim to have re-invented Ulam-Teller could only be partially true since Penney knew that a two-stage device was used by the United States, probably as a result of his trips to Los Alamos. But it seems to me that Ward did do what was asked of him, namely that given the requirement of a two-stage device, he realised that the two stages had to be in a heavy metal container; that the Ulam-Teller concept was based on thermal equilibrium between matter and radiation and this allowed the matter to be extremely compressed by radiation, thereby providing the conditions for fusion to take place. Thus Ward re-invented radiation implosion at Aldermaston. That was his crucial contribution.

---

[6] This is known as a hohlraum configuration

Just as Ulam and Teller had done before him, Ward realised that to make progress, Fermi's ideas for the Super had to be discarded. Roberts may have helped him but given Ward's characteristic intensely personal working habits, it is much more likely that Ward had the ideas both of thermal equilibrium and extreme compression and asked Roberts to do various associated calculations such as the calculation on uranium opacity referred to in his memoirs. Roberts then reported on their work just after Ward had left Aldermaston.

Richard Moore, Kate Pyne's successor at Aldermaston told me that "It [Ward's contribution] was reviewed at length by official historian Lorna Arnold, who concluded firmly that Ward's ideas were "not the basis of the British H-bomb". Her successor Kate Pyne reviewed the evidence available to Arnold again in a later study, which suggested that Ward might "possibly" have contributed one of the key concepts, radiation implosion, to the design process. However, others were also working on these concepts and, when Ward left AWRE, the path to Britain's H-bomb was still not yet clear. The weight of evidence from all of the relevant minutes, papers and drawings led Pyne to conclude that, far from relying on a single dramatic insight, H-bomb science was a fundamentally collaborative process involving a team of people working on a wide range of ideas and calculations."

Arnold, Pyne and Moore are not physicists. In the case of the H-bomb the original theory of the Super as enunciated by Fermi was that thermal equilibrium of matter and radiation was not possible and that the ignition temperature did not depend on compression. Ulam and Teller were working separately on the problem and then collaborated on a paper. Ulam had the idea of a fission primary initiating a fusion secondary in a heavy metal container; Teller the idea of thermal equilibrium which led to radiation implosion. Their work constituted the basic physics of Mike and subsequent US H-bombs . But Ulam and Teller were not the designers of those H-bombs. Those were collaborative efforts.

Ulam and Teller threw out Fermi's ideas and discovered that ignition was possible in thermal equilibrium at high compression. In the UK Ward[7] four years later did the same. As he says, "the discovery of radiation implosion cannot be assigned to more than one person" (Appendix II(v). The H-bombs tested in the Grapple series in 1957 and 1958 were based on that concept.

Arnold concluded "that the weapon concept that Ward had produced in 1955 was not developed " and was not the basis of the British H-bomb". According to Moore Pyne concluded that, far from relying on a single dramatic insight, H-bomb science was a fundamentally collaborative process involving a team of people working on a wide range of ideas and calculations."

I do not agree with these conclusions. I hope that I have demonstrated that Ward did have a 'single dramatic insight', namely that matter and radiation could be in thermal equilibrium and that in that event radiation pressure was proportional to the fourth power of temperature . That in turn meant that radiation implosion of the secondary was the dominant process in its detonation. That there was a subsequent "collaborative process involving a team of people working on a wide range of ideas and calculations." is not disputed; nor did Ward design the British H-bombs tested in the Grapple series, The principal designers of those were Keith Roberts and Bryan Taylor, just as the principal designer of Mike was Richard Garwin (Broad 2001).

But both the British and US H-bombs depended on thermal equilibrium, radiation implosion and high compression.. Ward was responsible for introducing those concepts into the British programme, just as Ulam and Teller were for US weapons.

---

[7] Ward told me that Penney gave him information on the Super which Penney kept in his safe. That must have included Moon's notes on Fermi's lectures and probably Fuchs' notes as well.

## 9) Epilogue

John Ward ended up as an embittered man. Dyson told me that "he became obsessed with the lack of recognition of his achievements. At the end of his life he was a tragic figure, isolated by his own querulous complaints[8]"

His Macquarie pension may not have been bad but he would not have got much income from his short periods at American universities. Additionally he thought that he should be financially compensated for his work at Aldermaston He even got Salam to write to Mrs Thatcher (Appendix II) saying that " I strongly feel that, at the time of his need, Her Majesty's Government might make a monetary gesture by either giving him a supplement of his pension or by some suitable appointment where his scientific talents can still be used." Not surprisingly, the government refused. Salam also stated that he did not know the details but "having collaborated with Professor Ward myself and knowing his calibre in fundamental Particle Physics, I would believe that he did indeed reinvent the process which was subsequently used by the Aldermaston Laboratory in building the British nuclear deterrent. Professor Ward received no recognition for his work by the British Government."

Nor did he receive what he considered to be his due in QED and electroweak theory. Salam won the 1979 Nobel Prize with Glashow and Weinberg for the prediction of neutral weak currents, not Ward, although Ward collaborated with Salam on their 1964 paper predicting neutral currents and also their original 1959 paper on the subject. And although Sakharov (1989) gave him his due as one of the 'titans of modern physics' with Feynman, Schwinger, Tomonaga , Dyson and Wick for his work on QED and Dyson told me that "Ward and I had an approximately equal share in the evolution of QED into its modern shape", Schweber in his history of QED devotes 100 pages to Dyson and a single page reference to Ward.

---

[8] Frank Duarte, who spoke to Ward frequently towards the end of his life told me that "I agree with Freeman Dyson that John was bitter about his lack of recognition. However from the late 1970s to late 1990s time appears to have have done some healing and tempered that bitterness" (Duarte 2020).

Yet his achievements did receive recognition. He was elected a Fellow of the Royal Society in 1965; he received the Guthrie Medal of the Institute of Physics in 1981; the Dannie Heineman Prize of the American Physical Society in 1982 and the Hughes Medal of the Royal Society in 1983.

Ward's obsession with MI6 and paranoiac view of the British and Australian governments were shared by many in Australia in the 1960s and 1970s[9]. After all one year after Ward took up his post at Macquarie the Australian Prime Minister Harold Holt disappeared while swimming and his body was never recovered. In 1975 the Australian Prime Minister Gough Whitlam was removed from office by the Governor General for the first and only time, possibly because Whitlam opposed the United States new signals intelligence station at Pine Gap, near Alice Springs. Conspiracy theories multiplied. Then in 1976 the ex-MI5 agent Peter Wright retired to Tasmania where he wrote his memoirs called 'Spycatcher'. In the book he said that he had been Chairman of a joint MI5/MI6 committee whose job was to review Soviet penetration in the British Security Services. Wright alleged that Sir Roger Hollis, the Director-General of MI5 from 1956 to 1965 was a Soviet spy, as was Harold Wilson, the former British Prime Minister. [Sir Antony Blunt, the Surveyor of the Queen's Pictures really was a Soviet spy].

There was also the Bogle saga. Gilbert Bogle was a Rhodes Scholar from New Zealand and a contemporary of Ward at the Clarendon in Oxford: he received his D. Phil. in Physics in 1952. He then worked in Sydney in the new field of masers. His work was considered outstanding and he was offered a position in

---

[9] I don't include references to the events in Australia mentioned in this section. Those events are well-described in Wikipedia.

Quantum Electronics by Bell Telephone Research Laboratories in New Jersey. On 1 January 1963 Bogle's body was found on the banks of a river in Sydney. The cause of death was not established. No one was ever prosecuted. Peter Wright believed that Hollis had recommended Bogle to the Australian Security Intelligence Organisation. For many years Ward corresponded with Dalitz in Oxford. Much of the correspondence was about physics and which jobs might be available. Much was also about conspiracies involving intelligence agencies and Bogle. For example in his letter to Dalitz of November 16 1987, Ward wrote that a former member of the Australian intelligence agency had claimed that Holt had been murdered and had linked that murder with Bogle's "murder". He added that it was highly likely that "both the Australian Atomic Energy Commission and Australian Security Intelligence Organisation were largely controlled by the British " (JCW 4a) .

In his review of Monk's biography of Oppenheimer, Freeman Dyson says that Oppenheimer never made any revolutionary discoveries in science although he was capable of doing so. He was too interested in the mainstream and the fashionable (Dyson 2013). Ward was the opposite: he was neither interested in the mainstream nor what was fashionable: he spurned opportunities to become the Lucasian Professor at Cambridge and the Wykeham Professor at Oxford and positions at other prestigious universities, and settled down at Macquarie. Yet he did make three revolutionary advances in physics which unusually can be simply described by short equations. First we have we his wave equation for two entangled photons

$$|1, 2> = (|\alpha, \beta> - |\beta, \alpha>)(|k, -k> - |-k, k>) \qquad (1)$$

describing the two-photon wave function for $J = 0$ where the photons are going in opposite directions, which leads to the first known derivation of the expression for the polarisation correlation of the two photons in a quantum entanglement situation.

Second we have his identity

$$Z_1 = Z_2 \qquad (7)$$

which with its generalisations shows the deep connection between gauge invariance and renormalisation in modern quantum field theory.

Third we have the Stefan-Boltzmann expression for the radiation pressure at temperature T arising from the thermal equilibrium of radiation and matter

$$S = AT^4 \text{ W/cm}^2 \qquad (12)$$

which Ward showed led to radiation implosion at high temperature, hence providing the conditions for the thermonuclear fusion of deuterium and tritium nuclei.

John Ward married Sarah Levin in April 1966 and was subsequently divorced.

He died on 6 May 2000.

## 10) Acknowledgements

I should like to thank John Charap, Frank Duarte, Freeman Dyson, Roger Elliott, Peter Knight, Chris Llewellyn-Smith, Richard Moore, Kate Pyne, Adam Roberts, Martin Rees and Dmitri Vassiliev for their help.

(17). 1960 (with J. M. Luttinger) Ground- State Energy of a Many-Fermion System II, *Phys. Rev.* **118** 1417-1427

(18). 1960 (with A. Salam) $\Delta I=1/2$ Rule, Phys. Rev. Lett. **5** 390

(19). 1961 (with A. Salam) On a Gauge Theory of Elementary Interactions *Nuovo Cimento* **19** 165-170

(20). 1961 (with A. Salam) Vector field associated with the unitary theory of the Sakata model, *Nuovo Cimento* **20** 419–421

(21). 1961 (with A. Salam) On the Symplectic Symmetry, *Nuovo Cimento* **20** 1228-1230

(22). 1963 (with E. W. Montroll and R. B. Potts) Correlations and Spontaneous Magnetism of the Two-Dimensional Ising Model, *J. Math, Phys.* **4** 308-322

(23). 1964 (with A. Salam) Electromagnetic and Weak Interactions *Phys, Lett.* **13** 168-171

(24). 1964 (with A. Salam) Gauge Theory of Elementary Interactions, *Phys. Rev.* **136** B763-768

(25). 1978 General relativity, the Dirac equation, and higher symmetries, Proc. Natl. Acad. Sci. **75** 2568

(26). 1992 'Nunca me arrependido que fiz' ('I've never been sorry about what I did') by John Ward in *Publico* (Lisbon), 4 April 1992.

(27). 2004 Memoirs of a Theoretical Physicist, *Optics Journal*, Rochester N.Y. http://www.opticsjournal.com/JCWard.pdf

## Correspondence

The correspondence referred to in the text of Ward (JCW) with Dalitz (RHD) and Dombey (ND) is listed below.

(a) Correspondence between Ward and Dalitz

RHD1   Letter from RHD to JCW, 18 December 1986
RHD2   Letter from RHD to JCW, 22 December 1986
RHD3   Letter from RHD to JCW, 13 March 1987
RHD4   Letter from RHD to JCW, 6 July 1987

    RHD5    Letter from RHD to JCW, 25 August 1989
    JCW1a    Letter from JCW to RHD, 22 February 1982
    JCW2a    Letter from JCW to RHD, 15 May 1987
    JCW3a    Letter from JCW to RHD, 16 November 1987

(b) Correspondence between Ward and Dombey

ND1 Letter from ND to JCW, April 30 1991
JCW1b    Letter from JCW to ND, September 17 1991

## References to Other Authors

**Appendices**

Appendix I

(i)  Page 36 of Ward's Thesis

Appendix II

(i)  Letter from JCW to Mrs Thatcher, May 13  1985

(ii)  Enclosure dated May 11 1983 sent to Mrs Thatcher with letter (i)) above

(iii)  Letter from Abdus Salam to Mrs Thatcher dated 3 August 1985

(iv)  Letter from P. R. Guyett to Abdus Salam, 20 March 1986

(v)  Letter from JCW to P. R. Guyett, 23 April 1986

# Appendix I

**Lemma.**

$$\text{let } A_n(\sigma) = \int_1^\sigma \int_2^{\sigma'} \cdots \int_n^{\sigma^{n-1}} \mathcal{H}(x^1)\mathcal{H}(x^2)\cdots \mathcal{H}(x^n)\, dx^1\cdots dx^n$$

and $B$ be any operator.

Then $\bar{A}_r(\sigma) B + \bar{A}_{r-1}(\sigma) B A_1(\sigma) + \cdots + B A_r(\sigma)$

$$= (-1)^r \int_1^\sigma \int_2^{\sigma'} \cdots \int_r^{\sigma^{r-1}} [\mathcal{H}(x^1), [\mathcal{H}(x^2), \cdots [\mathcal{H}(x^r), B]\cdots]]\, dx^1\cdots dx^r$$

The proof is by induction. Suppose it true for $n < r$

then $\bar{A}_{r-1}(\sigma) B + \bar{A}_{r-2}(\sigma) B A_1(\sigma) + \cdots + B A_{r-1}(\sigma)$

$$= (-1)^{r-1} \int_1^\sigma \int_2^{\sigma'} \cdots \int_r^{\sigma^{r-2}} [\mathcal{H}(x^1), [\mathcal{H}(x^2), \cdots [\mathcal{H}(x^{r-1}), B]\cdots]]\, dx^1\cdots$$

Instead of $B$ put $[\mathcal{H}(x^r), B]$ where $x^r$ is on $\sigma'$

$\bar{A}_{r-1}(\sigma) [\mathcal{H}(x^r), B] + \bar{A}_{r-2}(\sigma) [\mathcal{H}(x^r), B] A_1(\sigma) + \cdots$

$$= (-1)^{r-1} \int_1^\sigma \int_2^{\sigma'} \cdots \int_r^{\sigma^{r-2}} [\mathcal{H}(x^1), [\cdots [\mathcal{H}(x^r), B]\cdots]]\, dx\cdots$$

This may be written as

$$\sum_{\delta\sigma(x^r)} \left\{ \bar{A}_r(\sigma) B + \bar{A}_{r-1}(\sigma) B A_1(\sigma) + \cdots \right\}$$

Integration with respect to $x^r$ over space-time up to $\sigma$ then completes the induction for $n = r$, and the case $n = 1$ is obvious.

Page 36 of Ward's Thesis

# Appendix II

(i)     Letter from JCW to Mrs Thatcher, May 13  1985

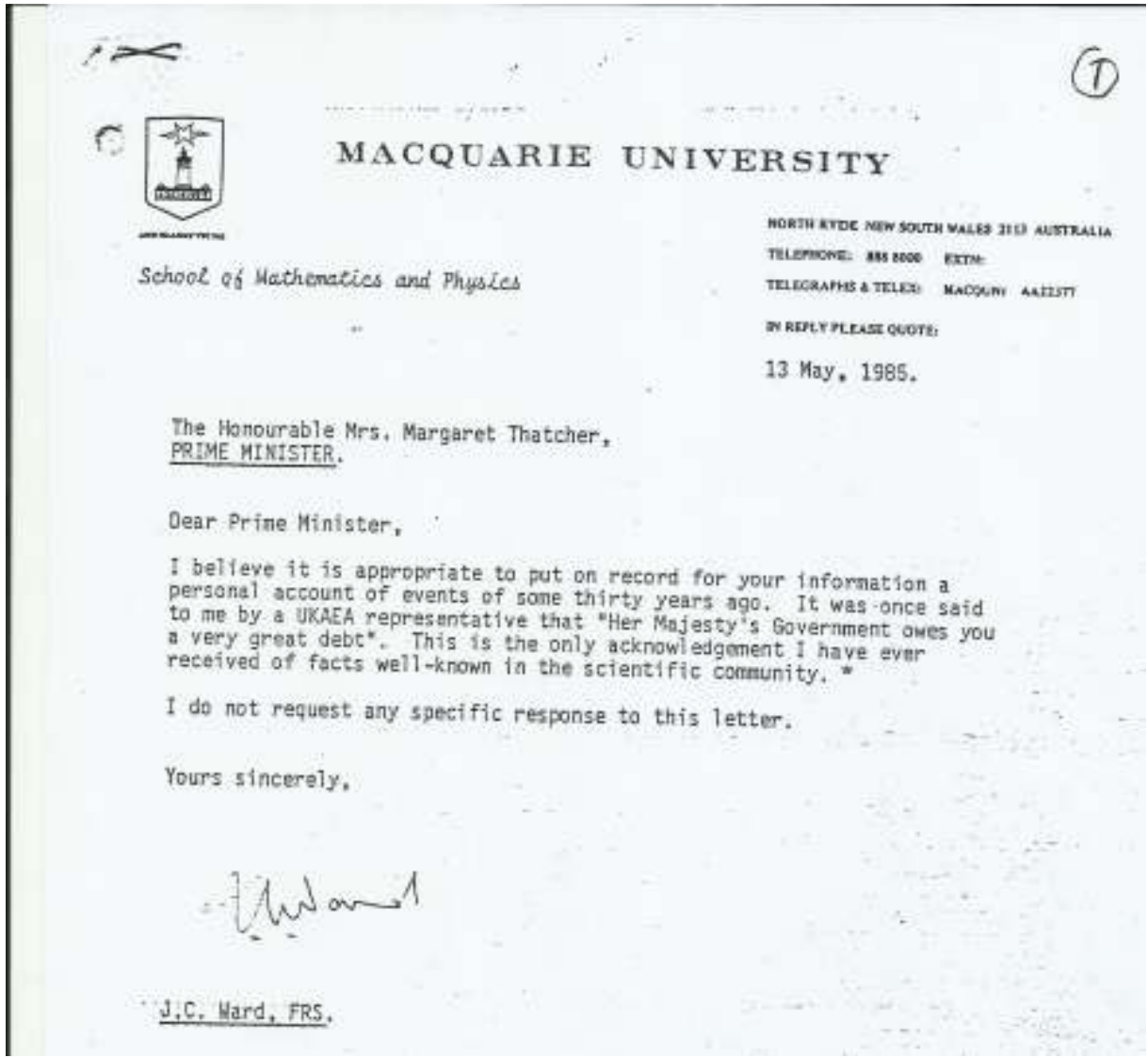

## ii) Enclosure dated May 11 1983

**1955**

In the spring of 1955 advertisements were prominently displayed for theoretical physicists to join the staff at Aldermaston at, by U.K. standards, quite attractive salaries.

Wishing to return to the U.K., and with a marriage in prospect, after further enquiries and negotiations, I was offered a position but decided to refuse, intending to take up either a possible position in Cambridge, or to return to the U.S.

When I telephoned William Cook to tell him of this decision, he was so upset that I said I would come if the matter was sufficiently urgent. He said it was indeed most urgent. Going to Aldermaston under these conditions clearly involved an extreme professional and personal risk and indeed as it turned out sacrifice. (The marriage did not take place.)

I now know why Cook was so anxious for me to come. He had seen a letter Kramers had written to Simon urging my return to Oxford in most determined language. Cherwell had used this letter improperly for his own purposes.

To my amazement when I reached Aldermaston, I was assigned the improbable job of uncovering the secret of the ULAM-TELLER invention, an idea of genius far beyond the talents of the personnel at Aldermaston, a fact well-known to both Cook and Penney.

Under great stress, with no assistance whatsoever, I came up with the correct scheme within six months, minor modifications excepted. When presented at a subsequent meeting, a crucial one judging by the full-dress uniform of the visiting Admiral, my proposal, the only one offered, was peremptorily rejected by Penney, who declared the matter not to be urgent anyway! I was supported barely pro-forma, if at all, by Cook. Afterwards Penney demonstrated his complete lack of understanding of the problem in a private talk with Cook and myself. I was not invited to subsequent meetings held to discuss the project.

I therefore quite correctly and naturally resigned forthwith, and returned to the U.S. taking the first job I could get. My personal and profession survival of this trauma was something of a miracle.

**1985**

Last year I retired at 60, believing it not too late to contribute my remaining talents to the recalcitrant problems of modern physics and expecting to be able to travel more widely than possible otherwise. I now find that I am afflicted with chronic high blood pressure. The prognosis is at the moment uncertain. Clearly this is a result of past stress, and I see no reason to exclude from this my Aldermaston experience.

SYDNEY, MAY 11, 1985.

### (iii) Letter from Abdus Salam to Mrs Thatcher dated 3 August 1985[10]

<div style="text-align: right;">3 August 1985</div>

Dear Prime Minister

In May this year, Professor John C. Ward, retired Professor at Macquarie University, Australia, wrote to you regarding his role in uncovering the secret of the ULAM-TELLER invention at Aldermaston. I do not know the details but having collaborated with Professor Ward myself and knowing his calibre in fundamental Particle Physics, I would believe that he did indeed reinvent the process which was subsequently used by the Aldermaston Laboratory in building the British nuclear deterrent. Professor Ward received no recognition for his work by the British Government.

After retirement from Macquarie, he finds he is in straitened circumstances with a chronic debilitating disease. As I said before, Professor Ward was my collaborator for part of the work for which Professors S. Glashow, S. Weinberg and I were awarded the Nobel Prize in Physics in 1979. This work concerns the unification of fundamental forces. I wrote two out of three (often quoted) papers on this subject with John Ward. In view of Professor Ward's calibre and in view of his undoubted services to the United Kingdom I strongly feel that, at the time of his need, Her Majesty's Government might make a monetary gesture by either giving him a supplement of his pension or by some suitable appointment where his scientific talents can still be used.

With my very best wishes

<div style="text-align: center;">
Yours sincerely

Abdus Salam

Professor of Theoretical Physics

Imperial College of Science and Technology
</div>

---

[10] The copy of this letter that I have is too faint for me to scan.

(iv) Letter from P. R. Guyett to Abdus Salam, 20 March 1986

From: Mr P R Guyett, Head of Civilian Management (Specialists)

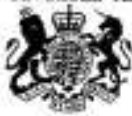

MINISTRY OF DEFENCE

St Christopher House, Southwark Street, London SE1

Telephone 01-921-

PERSONAL

Professor Abdus Salam FRS
Professor of Theoretical Physics
Department of Physics
Imperial College of Science and
 Technology
Prince Consort Road
London SW7

Our Reference:
D/CM(S)1/5/1/13
HD/P1

20 March 1986

Dear Professor Salam

**Professor John C Ward FRS**

You wrote to the Prime Minister in August last year to express your strong feelings that in view of Professor Ward's calibre and his undoubted services to the UK, a monetary gesture should be made to him, at a time of need, by either giving him a supplementation of his pension or by some suitable appointment where his scientific talents can still be used.

Please accept my apologies for the delay in replying to your letter, which has occurred whilst various necessary enquiries have been made.

May I also say that we are sorry to learn of Professor Ward's medical affliction, restricting his ability to travel and make the contributions that his widely respected talents would otherwise have enabled him to make.

On the central question of providing a monetary award to Professor Ward, our enquiries have shown that no such ex-gratia payments have been made to any of the staff involved in the early H-bomb work, or any related programmes undertaken by the UKAEA over that period. As you know, it is more customary for civil servants to receive honours for exceptional services rather than additional payments, other than those that accompany promotion to more senior ranks.

Despite the policy background, my colleagues at AWRE have made very careful enquiries into the events of 30 years ago in order to establish whether Professor Ward's contribution to the programmes at that time would provide the grounds for any exceptional application for an ex-gratia payment.

A check back over that length of time is of course difficult to accomplish. In addition to consulting the technical and other records that are available, advice has been obtained from many of the key people of the time, nearly all of whom are now retired.

(iv)   Letter from P. R. Guyett to Abdus Salam, page 2

The outcome of all this leaves great room for uncertainty. Professor Ward's employment covered some six months up to January 1956, at the grade level of SPSO. Given his age, the grade level itself shows how very well regarded he was. It is clear also, that there was considerable disappointment that he chose to leave after such a short period. Despite the extent of the enquiries, however, we lack the evidence to unravel Professor Ward's contribution from amongst those of the other high calibre people then at AWRE, during part of the critical stage in the UK H-bomb development.

Our answer must therefore be that we regret we cannot justify any payment in the circumstance that there is no relevant precedent for such a payment or the evidence that would enable us to treat the application as an exception. This is in no sense a reflection on Professor Ward's scientific ability.

I regret also that we have been unable to identify a consultative task that would be appropriate in Professor Ward's present situation.

My apologies, once again, for the time it has taken to resolve this matter.

Yours sincerely

PR Guyett

(v)   Letter from JCW to P. R. Guyett, 23 April 1986

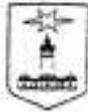

# MACQUARIE UNIVERSITY
SCHOOL OF MATHEMATICS & PHYSICS

NORTH RYDE, NEW SOUTH WALES 2113 AUSTRALIA
TELEPHONE: 888 8000   EXTN: 9050
TELEGRAPHS & TELEX: MACUNIV AA22377

Refer: Your reference
       D/CM(S)1/5/1/13
       HD/P1

IN REPLY PLEASE QUOTE: JCW/gla

23rd April, 1986

Mr. P.R. Guyett,
Head of Civilian Management (Specialists),
MINISTRY OF DEFENCE,
St. Christopher House,
Southwark Street,
LONDON SE1   U.K.

Dear Mr. Guyett,

Professor Salam has sent me a copy of your letter (dated 20 March, 1986).

It seems to me that you may not have had proper access to classified information. The discovery of the principle of radiation implosion cannot be assigned to more than one person, and the quote from Robert Jastrow, (see enclosure), coming from the CIA evidently, confirms this, as does the phrase quoted from the former commercial director of British Nuclear Fuels.

Apparently Professor Salam suggested a pension. This was not the point of my letter to Mrs. Thatcher. That proper credit should be given now seems to me quite clear, simply as a matter of record. The attempted obfuscation of this history by the U.K. authorities has had most serious consequences, particularly in Australia.

Sincerely,

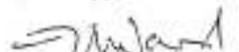

PROFESSOR J.C. WARD

*This is explained in ⑥ (i.e. the Bogle Affair)*

Copy to Prof. A. Salam

Enclosure 1